# Algorithmically Curated Lies:
## How Search Engines Handle Misinformation about US Biolabs in Ukraine

Elizaveta Kuznetsova[a], Mykola Makhortykh[b], Maryna Sydorova[b],
Aleksandra Urman[c], Ilaria Vitulano[a], Martha Stolze[a]

*Abstract*. The growing volume of online content prompts the need for adopting algorithmic systems of information curation. These systems range from web search engines to recommender systems and are integral for helping users stay informed about important societal developments. However, unlike journalistic editing the *algorithmic information curation systems (AICSs)* are known to be subject to different forms of malperformance which make them vulnerable to possible manipulation. The risk of manipulation is particularly prominent in the case when AICSs have to deal with information about false claims that underpin propaganda campaigns of authoritarian regimes. Using as a case study of the Russian disinformation campaign concerning the US biolabs in Ukraine, we investigate how one of the most commonly used forms of AICSs - i.e. web search engines - curate misinformation-related content. For this aim, we conduct virtual agent-based algorithm audits of Google, Bing, and Yandex search outputs in June 2022. Our findings highlight the troubling performance of search engines. Even though some search engines, like Google, were less likely to return misinformation results, across all languages and locations, the three search engines still mentioned or promoted a considerable share of false content (33% on Google; 44% on Bing, and 70% on Yandex). We also find significant disparities in misinformation exposure based on the language of search, with all search engines presenting a higher number of false stories in Russian. Location matters as well with users from Germany being more likely to be exposed to search results promoting false information. These observations stress the possibility of AICSs being vulnerable to manipulation, in particular in the case of the unfolding propaganda campaigns, and stress the importance of monitoring these systems' performance to prevent it.

[a] Weizenbaum Institute for the Networked Society, Berlin, Germany,
elizaveta.kuznetsova@weizenbaum-institut.de
[b] Institute of Communication and Media Studies, University of Bern, Bern, Switzerland
[c] Department of Informatics, University of Zurich, Zurich, Switzerland

# 1. Introduction

Recent discussions about the online news ecosystem are often triggered by crises, from interference of rogue political actors into the affairs of democratic states (Mueller, 2019) and the increase of misinformation campaigns globally (Muhammed & Mathew, 2022) to manipulation of public opinion through wartime propaganda (Litvinenko, 2022). To manage the abundance of online content and to keep online news environments in check, digital platforms such as Google and Meta have been introducing content filtering and ranking mechanisms, powered by *algorithmic information curation systems (AICSs)*. Aimed at "organizing, selecting and presenting subsets of a corpus of information" to users (Rader & Gray, 2015), AICSs have made platforms de-facto gatekeepers of online information, significantly affecting how individuals perceive socio-political phenomena (DeVito, 2016; Nechushtai & Lewis, 2019).

The power of AICSs to define the global information ecosystem has raised concerns over possible forms of system malperformance and its potential consequences. Specifically, studies have demonstrated the risk that AICSs can amplify stereotypes and potentially contribute to discrimination, in particular of vulnerable groups (Noble, 2018; Ottenbacher et al., 2017; Urman & Makhortykh, 2022). Manipulation of AICS by foreign governments or domestic political actors has been another area of concern. Some of these manipulations fall in the realm of computational propaganda, where the platforms' AICSs are used to facilitate and amplify the spread of false content (Bradshaw, 2019; Stukal, Sanovich, & Tucker, 2017; S. C. Woolley & Howard, 2016).

Relevant research is extensive and primarily focuses on the role of bots in viral information distribution (Crothers et al., 2019; Im et al., 2020; Zannettou et al., 2019) or on developing tools to prevent the spread of it by identifying and removing false claims (Aguerri & Santisteban, 2022; Garon, 2022; Saurwein & Spencer-Smith, 2020). Only recently has scholarship started investigating the role of AICSs in the distribution of false information and propaganda (Bradshaw, 2019; Hussein et al., 2020; Kuznetsova & Makhortykh, 2023; Makhortykh & Bastian, 2022; Srba et al., 2023; Toepfl et al., 2022). Despite burgeoning literature in the field, however, there is still limited understanding of how different factors that AICSs take into consideration (e.g. input language or user location) affect platforms' ability to deal with false information. Particularly rare are cross-language studies on AICSs with a focus on non-Western languages (Urman et al., 2024). We also do not fully know how different forms of AICSs' malfunctioning—for instance, random errors related to prioritization of non-relevant content or systematic bias related to disproportionate visibility of specific viewpoints— mean for algorithmic handling of false information. The lack of knowledge on AICSs of web search engines is problematic due to the fact that digital platforms have for long acted as gatekeepers of online information (Nielsen, 2016) and have the power to sway opinions (Epstein & Robertson, 2015). Understanding how search engines manage false content is thus of particular importance amid the ongoing war in Ukraine as the use of computational propaganda for manipulating the public opinion by Russia has substantially intensified (Canada, 2022; Pasi, 2022) and therefore manipulation of curation algorithms on search engines may likely be part of it.

To achieve better understanding of the possible manipulation of web search engines in the context of ongoing propaganda campaigns, we look at how a selection of major search engines deals with information about false claims referring to US-funded biolabs in Ukraine. This pro-Kremlin narrative, aimed at justifying Russian aggression against Ukraine, has found an unusually large audience and has been feeding into the existing conspiracy theories popular among the far-right community in the US (Chappell & Yousef, 2022). While false claims relating to biological warfare have a long tradition in Russian propaganda since the Cold War (Boghardt, 2009; Chappell & Yousef, 2022), the promotion of this particular false narrative is instrumental to the Kremlin due to its ability to construct Ukraine and the West as an existential threat to Russia. The twofold function of this ideational weaponry can be seen as, on the one hand, mobilizing internal support of the invasion within the Russian population, and reducing the support for Ukraine in the West, on the other.

In this research we use agent-based algorithmic auditing to investigate what role the search engines Google, Bing, and Yandex play in the distribution of the 'US biolabs' misinformation narrative and pose the following research questions:

RQ1. What information sources do different search engines prioritize in response to the 'Ukraine biolabs' search query?
RQ2. How does the selection of these sources vary depending on the search engine used and the language of the query?
RQ3. Do these sources mention, promote or debunk misinformation and does the exposure to false information vary depending on the user location and whether it changes over time?

The paper proceeds as follows: after a literature review and presentation of our methodology, we present a descriptive analysis of our data, supplemented with the multinomial regression analysis, and a discussion of our results. This study advances our knowledge on the role of search engines in managing online information environments and discusses potential avenues for improving the online news ecosystems.

## 2. Organizing Information Environments Online

### 2.1. Algorithmic Information Curation

AICSs are a set of automated decision-making practices used by online platforms to organize content and provide users with a "usable and coherent package" of information (Polgar, 2021). These practices are essential for tackling information overload caused by the exponential increase in the volume of online content (Rodriguez et al., 2014). There are different forms of AICSs which vary depending on their function within the platform. One of them is *algorithmic content moderation*, designed to filter out content that violates platform rules and can be damaging for individuals or groups (Gorwa et al., 2020). Besides these platform-driven AICSs, there is also a growing number of user-made algorithmic curation mechanisms, for instance, which often rely on robotic agents (or bots) deployed by users within a specific platform used to prevent vandalism or propagation of incivil content, for example on Wikipedia or Discord (Makhortykh et al., 2022). Another form of AICSs is content retrieval systems such as recommender systems (Ricci et al., 2015) or *search engines* (Lewandowsky, 2023), which is the focus of our article.

While these mechanisms of content retrieval on search engines are not always personalized, the development of AI increasingly allows individualizing user experience by exposing them to more tailored content (Berman & Katona, 2016). Specifically, AICSs customize information based on user characteristics and, more recently, adopt conversational interfaces powered by AI to enable information retrieval in a dialogic format (Gude, 2023). Such personalization can be driven by different factors from users' previous patterns of actions (Feuz et al., 2011; Polgar, 2021) to user location (Kliman-Silver et al., 2015; Kuznetsova & Makhortykh, 2023) and language of user input in the retrieval system (Urman, Makhortykh, & Ulloa, 2022). Personalization of content retrieval enables individuals to be informed about important social events and developments as well as can potentially nudge users towards expanding their information diets. However, it also raises a number of concerns regarding the potential risks of personalization amplifying inequalities in terms of information access between different groups of users and contributing to societal polarization (Jaeho Cho & Luu, 2020). Other concerns include algorithmic content retrieval systems discriminating against certain groups by making content produced by them or about them less accessible (D. Lazer et al., 2023).

The multiplicity of factors influencing the functionality of AICSs on search engines makes it challenging to study these systems. Algorithmic curation has been described by Daucé and Loveluck (2021) as an "invisible hand", whose rules are unintelligible and that decides which information and news outlets will be deemed as relevant and presented to users (Daucé & Loveluck, 2021). It is due to its lack of transparency that algorithmic content retrieval has received burgeoning scholarly attention, particularly, for its potential negative consequences, namely reinforcing of biases and feedback loops (Rader & Gray, 2015).

This lack of transparency and the potential for biased performance amplifies the risks of automated content retrieval systems being instrumentalized for computational propaganda and spreading misinformation. Despite the substantial amount of research focused on possible negative consequences of malfunctioning of these systems, in particular search engines and recommender systems (e.g. Bandy, 2021), little is known about these systems' roles in this particular context. However, a few exceptions which we discuss in more detail below, suggest that studying the potential of automated content retrieval for countering or amplifying the spread of misinformation is of paramount importance.

### 2.2. Computational Propaganda on Platforms

Computational propaganda, the process involving a range of automated methods, such as bots and algorithms, to influence public opinion, has been a persistent concern in democratic and authoritarian political contexts (S. Woolley & Howard, 2017). Much of the current research on the phenomenon has been focused on false information (Nerino, 2021). Studies have mostly investigated data collected from Twitter (Alieva et al., 2022; Eady et al., 2023; Freelon & Lokot, 2020; Jang et al., 2018; Jarynowski, 2022; Stukal, Sanovich, & Tucker, 2017) and Facebook (Franklin et al., 2020). Methodologically, existing research uses approaches varying from manual and automated content analysis (S. Woolley & Howard, 2017) to network analysis (Alieva et al., 2022; Freelon & Lokot, 2020; Jarynowski, 2022) and survey research (Eady et al., 2023). The main focus of these studies has been on patterns of false information distribution and the reach of inaccurate claims online. For instance, Ghosh and Scott (2018) emphasized how Google, Facebook and Twitter, being at the center of a vast ecosystem of services and being able to deliver personalized content, represent a threat to democracy if being leveraged for propaganda and misinformation.

Concerning false information effects, much research has focused on election interference, with studies examining both the US (Muhammed & Mathew, 2022) and other democratic countries (D. M. J. Lazer et al., 2018). Moreover, studies have identified that misinformation spread is not only technology-driven, but is also shaped by national information environments (Humprecht, 2019; Humprecht et al., 2023) and is subject to a confirmation bias in a sense of exposure to specific misinformation content aligning with one's pre-existing beliefs (Cerf, 2017; Muhammed & Mathew, 2022). Jang et al. (2018), for example, find that fake news tweets mostly originate from ordinary users and undergo a process of transformation as they spread among users - and mostly spread among people with similar ideologies (Jang et al., 2018).

Russian disinformation campaigns have been a special focus of scholarly and political attention, with studies looking at the Kremlin's covert communication strategies (Fedor & Fredheim, 2017). Previous research on such computational propaganda has, however, primarily focused on detecting Russian bots and trolls by describing their behaviors and characteristics (Crothers et al., 2019; Grimme et al., 2017; Im et al., 2020; Stukal, Sanovich, Bonneau, et al., 2017; Zannettou et al., 2019). Moreover, scholars have focused on the most prominent actors of Russian propaganda, like RT (formerly, Russia Today) and RIA Novosty, as well as the more covert sources and online propaganda accounts (Franklin et al., 2020; Golovchenko et al., 2020; Grigoriy, 2022; Kuznetsova, 2021; Orttung & Nelson, 2019). While the body of literature on Russia's digital propaganda is growing, only a few studies to date investigate the role of algorithmic content retrieval in the distribution of misinformation, particularly (Kuznetsova & Makhortykh, 2023; Toepfl et al., 2022). This paper seeks to further develop this body of research.

## 2.3. Misinformation in Digital Search

Internet users tend to regularly verify facts they encounter online, considering information from search engines to be more trustworthy than traditional news sources. Especially young users have been found to trust search engines (Trevisan et al., 2018). Evaluating the accuracy of information through online search has also been a central element of many media literacy programs (Persily & Tucker, 2020). Indeed, it has been suggested that verifying accuracy of information by performing an online search can reduce misinformation effects (Hasanain & Elsayed, 2022). Recently, however, these beliefs have been disproven. Aslett et al. (2024) have demonstrated that using online search to assess the credibility of fake news articles can actually lead to a higher likelihood of believing these stories. Moreover, different search engines can construct rather different interpretations of social reality and, in turn, deceive their users (Jiang, 2013). In particular, politics-related issues can be represented in skewed ways due to search engine algorithms often unequally treating information about specific social groups or topics (e.g. (Epstein & Robertson, 2015; Kravets & Toepfl, 2021; Makhortykh et al., 2020; Steiner et al., 2022; Unkel & Haim, 2021).

Despite relative academic awareness about algorithmic bias, only a handful of existing studies have investigated the presence of false information on search engines and its spread. Looking at Google, Bradshaw (2019) focuses on SEO keyword strategies of junk news domains and finds that they are mainly driven by navigational searches (Bradshaw, 2019). Despite Google's efforts to curb the discoverability of these domains, disinformation producers can "find new ways to optimize their content for higher research rankings" (Bradshaw, 2019, p. 16). Recent work has also found evidence that pro-Kremlin pseudo think tanks have been artificially boosted in Google's search results using a network of low-quality websites, highlighting the potential for search engine manipulation by the authoritarian regimes in general and the Russian state in particular (Williams & Carley, 2023). Several recent studies have shown the potential weaponization of search engines based in authoritarian countries to spread specific narratives about politically important events. Focusing on reference and source bias, Kravets & Toepfl (2021) find that Yandex, the largest Russian search engine, favors pro-Russian-regime sources and potentially censors information about protest activities. By conducting an audit of the Yandex.News, a news aggregator service offered by Yandex, Daucè and Loveluck (2021) found that it is subject to tight control by Russian authorities and does not reflect the pluralism of news content circulating in the Russian digital space.

Contrasting findings have been presented by Metaxa-Kakavouli & Torres-Echeverry (2017) who find that in many cases Google is, in fact, not susceptible to political misinformation. Only 1.5% of the results found in the top ten outputs about U.S. congressional candidates to the 2016 presidential election came from websites that were described as "fake news domains" (Metaxa-Kakavouli & Torres-Echeverry, 2017). Similarly, Urman et al. (2022), found substantial variation in the number of conspiracy-promoting results in response to different conspiracy-related queries across multiple search engines with Google including the least number of such results. However, similar research using comparative algorithmic auditing to study the spread of Russian disinformation on various search engines, differentiating by location, language and time, is still lacking.

## 3.     Methods & Data Analysis

This study employed a mixed-method approach to investigate the performance of three search engines Google, Bing, and Yandex in response to misinformation-related queries in different languages. In designing this comparative study, the selection of search engines was strategically based on their use in different regions. Google and Bing are the two most commonly used search engines in the majority of Western countries, whereas Yandex is the largest Russian search engine and the one which is presumably influenced by the Kremlin to facilitate its disinformation campaigns (Makhortykh et al., 2022; Whalen et al., 2022).

Specifically, for our investigation we used virtual agent-based auditing, a research method that simulates human behavior for generating inputs for an algorithmic system, and then analyzed data collected in the course of an audit using qualitative content analysis for sources and content of search results in combination with multinomial regression analysis. In our case, algorithm audit involved the process of entering search queries and then retrieving system outputs, i.e. HTML pages of web search results. In contrast to other algorithm audit approaches (Bandy, 2021), virtual agent-based audits enable the investigation of the performance of the system in a controlled environment (Ulloa et al., 2022) that is important for understanding how different factors (e.g. location, time, or language) can influence the outputs which the system produces.

To implement the virtual agent-based audit, we built a cloud-based infrastructure using Google Compute Engine. The infrastructure consisted of the network of Debian-based virtual machines created from scratch. We then installed two internet browsers - Chrome and Firefox - on each virtual machine. We deployed one virtual agent per browser, modeled via Selenium, a collection of tools for browser automation. All resulted virtual agents performed a simple routine: opening the browser, entering the research query in one of the four languages of this study (English, German, Ukrainian, and Russian) in the text search of one of three search engines (using .com version for consistency), saving HTML files of the first page of search results, then closing the browser to remove cookies and browser history to prevent them from affecting the subsequent searchers.

The design of the audit reflected our interest in the different factors which can be taken into consideration by algorithmic information curation systems for personalizing their outputs. Firstly, we were interested in the impact of location from where the search was conducted. For this, we deployed our agents via three regional clusters of Google Compute Engine: Paris, Frankfurt, and Zurich, ensuring that our agents have French, German, and Swiss IP addresses. Secondly, we set out to track the difference between languages. We initially formulated our search query in English, using Google Trends tool, which resulted in the following search term: *ukraine biolabs*. We then translated this term into German, Ukrainian, and Russian. Lastly, we were interested in examining the impact of time when search was conducted. We, therefore, collected data in four rounds - in the morning and evening of June 21 and June 22, 2022 - aiming to see if there are differences over short periods of time considering the developing nature of misinformation narrative. Altogether, we deployed 42 virtual machines which resulted in 84 agents which were split between search engines and different locations in the following way: 10 agents per search engine for Frankfurt and Zurich and 8 agents per search engine for Paris.

For our analysis, we focused on the top 10 organic search results which were composed of 381 unique URLs out of 15,957 search results[1]. To analyze unique URLs, two coders conducted preliminary rounds of qualitative analysis of the data and designed a codebook, followed by three rounds of coding and control coding. Firstly, the coders identified the type of source (news site, blog, etc.) based on the "about" page of the site. Secondly, coders determined political leaning of the source to see whether retrieved sources are affiliated with political ideology or an authoritarian government (Russian or Chinese state), using the about page, secondary literature to identify funding sources, as well as the Media Bias/Fact Check website. Thirdly, the coders analyzed whether the articles contained conspiratorial content (yes/no) or misinformation (debunked/promoted/mentioned)[2].

---

[1] The uneven number of results is due to occasional variation in the number of organic search results per page.

The main piece of misinformation we encountered in the data concerned the false and prominently debunked story (Alieva et al., 2022) that the US had been funding biolaboratories in Ukraine that were either researching Covid-19; and/or developing biological weapons; and/or nuclear weapons. We use the term misinformation in this study to refer to false information for simplicity, given that our study has not per se focused on identifying the intent behind published information (Søe, 2021). Specifically, the coders identified whether the articles *mentioned*, *promoted*, or *debunked* misinformation, for example, by providing an explanation of why this story was misleading. The *mentioned* category included instances when information was false but it was unclear whether the source considered this information as true or not. Even though we introduce this level of nuance into the coding, we nevertheless consider all cases of not explicitly debunking the biolabs story as containing misinformation.

To test intercoder-reliability, we performed the Brennan-Prediger Kappa test which showed a high level of agreement for all categories: Misinformation [0.85], Source Type [0.88], Source Political Leaning [0.85], Conspiracy [0.89]. We consensus-coded the few disagreements we found, merged the coded links with the original dataset and performed several rounds of descriptive analysis, as presented below.

Additionally, we performed a multinomial logistic regression analysis to assess the association between exposure to misinformation and the characteristics of the queries. For this analysis, we excluded queries that did not mention biolabs-related facts, and used the command "multinom" of the "nnet" R package. In the regression, the dependent variable was "misinformation exposure" with three levels (Debunked, Mentioned, Promoted), while the independent variables were the time in which the query was performed (21.06.2022 morning, 21.06.2022 evening, 22.06.2022 morning, 22.06.2022 evening), language (English, German, Ukrainian, Russian), region (Switzerland, France, Germany), and search engine (Bing, Yandex, Google). We estimated odds ratios as well as 95% confidence intervals. Effect estimates whose 95% confidence interval included the null value were considered not statistically significant.

## 4. Results

### 4.1. Prioritization of information sources

Analyzing differences in types of sources returned by search engines (Figure 1), we find that for all languages, with a slight variation, all search engines prioritized news outlets as the source of information: between 54.4% and 89.7% of sources returned were news outlets with diverse political leanings and affiliations, like Bloomberg, Fox News or Russia Today, except for Google in Russian (47.2%) and Google in Ukrainian (44.9%). Google prioritized fact-checkers more than the other two search engines, but still a rather low level compared to the other source types. Specifically, Google featured between 7.1% to 10.0% of sources identified as fact-checkers across all languages. Bing displayed links to fact-checking websites for English in 10.1% of cases, followed by 8.6% in Russian, 0.4% in German, and 0.1% in Ukrainian.

---

[2] The codebook is available for review in Supplementary File #2

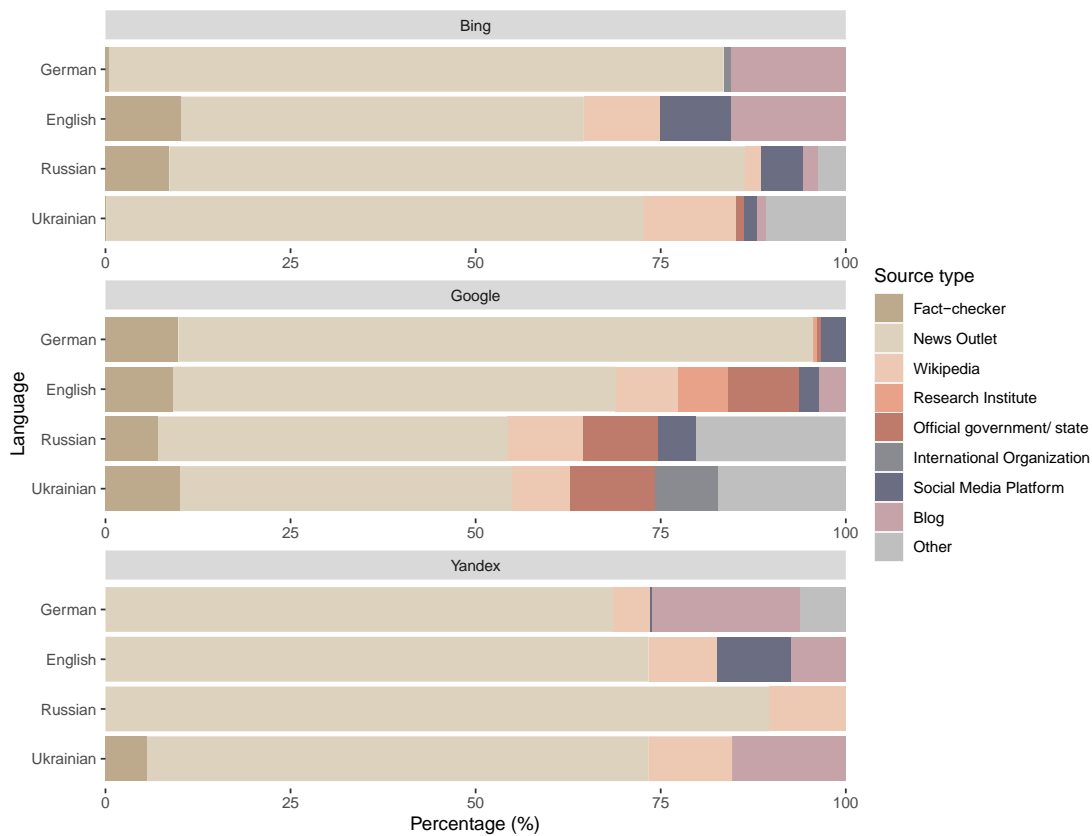

*Figure 1. Distribution of content by source type, language of the query, and search engine.*

The search queries on Yandex did not yield any fact-checking sources for English, German and Russian, which is in stark contrast to Yandex in Ukrainian, where the presence of fact-checkers is comparatively high (5.6%), for instance featuring articles by Detector Media or Vox Ukraine. This might be explained by the overall higher presence of Ukrainian fact-checkers, or a lack of curation by Yandex due to a lack of knowledge of Ukrainian. Interestingly, only Google displayed sources from official government sites, e.g. from the US embassy in Ukraine, for all languages (0.5% to 11.5% of search results). Bing and Yandex did not include any official government websites in the search results, except for a small fraction of results on Bing in Ukrainian (1.0%). Output by research institutes like Brookings only appeared in Google queries performed in English language (6.7%) and in German language (0.5%).

Search outputs from social media platforms like YouTube were only found on Bing in English (9.6%), Russian (5.6%) and Ukrainian (1.8%); on Google in Russian (5.1%), German (3.4%) and English (2.6%); and on Yandex in English (10.0%), and German (0.2%). Substantial variation was also observed in the case of blogs, linking to e.g. Anti-Spiegel, Strange Sounds or tkp.at. While appearing infrequently on Google (only 3.7% of sources in English), blogs were often present on Bing and Yandex. Around 16% in English and German Bing search results led to blogs, while only 1.3% and 2.1% for Ukrainian and Russian Bing, respectively. On Yandex, 20.1% of German and 15.3% of Ukrainian sources were blogs, but only 7.4% for English and none for Russian.

## 4.2. Distribution of information by political leaning of the source

Investigating the distribution of content by political leaning of the source (Figure 2), we see noticeable distinctions depending on the language of search. Russian state-sponsored media sources, for example, are present in all search engines. Unsurprisingly, Yandex presented the largest share of Russian state-sponsored results, with percentages ranging from 19.8% to 89.5% across languages. In particular, in Russian, Kremlin-sponsored sources were represented the most on Yandex (89.5%). Yandex search outputs in Ukrainian also displayed a number of Russian state-sponsored media (38.4%), even though it also featured a comparatively high presence of Western mainstream content compared to the outputs for the Russian query. Interestingly, even Bing and Google outputs presented some amount of Russian state-sponsored sources. For example, 42.9% of Google outputs for the Russian query were made of Kremlin-sponsored sources, whereas outputs for the Ukrainian query contained only 1.5% of such sources. Bing returned higher shares of Russian state-sponsored sources, especially for the Russian query (55.4%), but also for the English and German ones (15.4% and 16.9% respectively), and substantially smaller numbers for the Ukrainian one (0.8%).

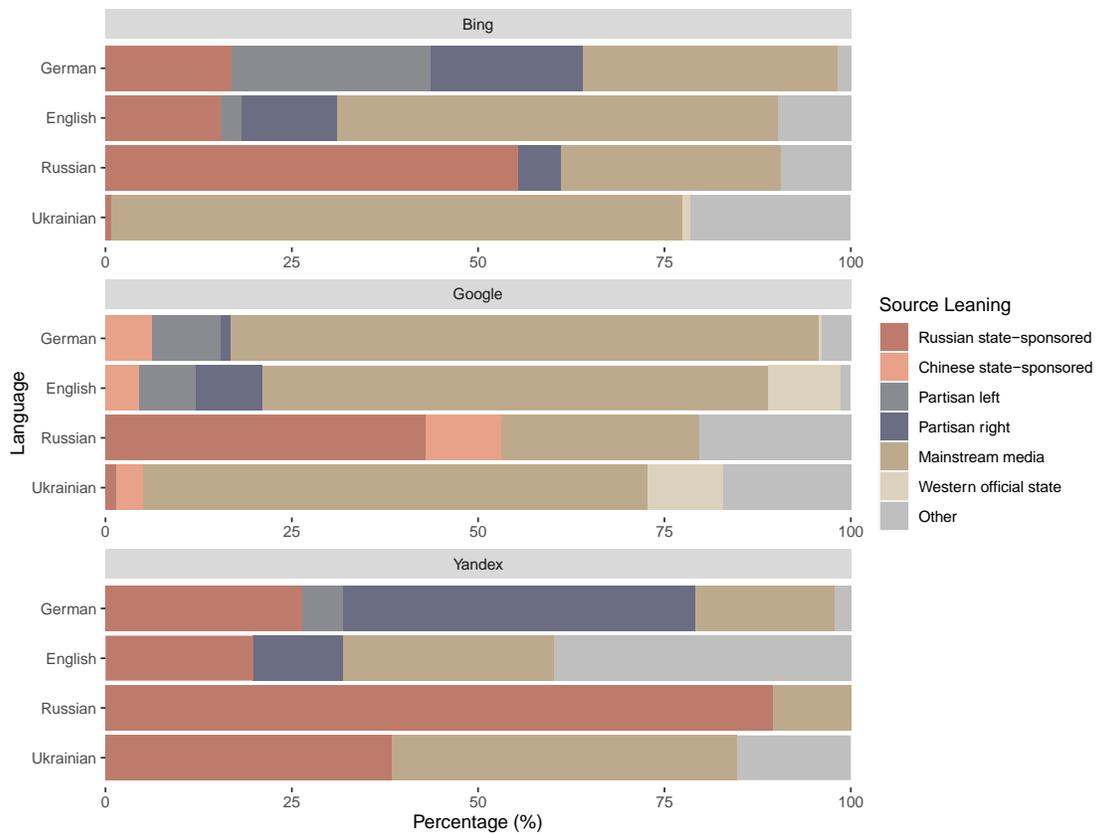

*Figure 2. Distribution of content by political leaning of the source, language of the query, and search engine.*

Chinese state sponsored results were present only on Google in all languages, with percentages ranging from 3.5% for Ukrainian to 10.1% for Russian. For partisan left- and right-leaning media, across all search engines, Bing in German language presented the largest share of left-leaning sources (26.7%), while the share of partisan left media on Yandex and Bing was very low (7.7% for English and 9.2% for German on Google, and reaching 5.5% on Yandex in German). In contrast, we found a much higher share of right-wing sources on all three search engines, with the two largest shares recorded on German Yandex (47.1%) and German Bing (20.5%).

Mainstream media represented the largest share of sources for all languages on Google (between 67.8% and 78.9%) except for Russian language (26.7%), where the largest share is represented by Russian state-sponsored results (42.9%). Mainstream media also represent the largest share of sources for Bing in Ukrainian (76.6%), German (34.1%), and English (59.1%). On Yandex, the share of mainstream media sources was substantially lower than the one observed on Google, with the exception of Yandex Ukrainian where mainstream media represented the main type of sources (46.2%).

### 4.2. Misinformation distribution on search engines

Overall, the results promoting misinformation were predominantly stemming from news outlets (69%). Figure 3 indicates the presence of misinformation depending on the search engine used and the language of the query. Among search engines, Google shows the largest shares of articles debunking the biolabs story (German 84.3%, English 76.9%, Ukrainian 77.8%), except for the Russian query, for which the share of such articles amounted to only 26.7%. The share of the articles promoting the misinformation was also lowest on Google (German 10.0%, English 19.9%, Ukrainian 22.2%), with the exception of the outputs for the Russian query where, among all combinations of search engines and languages, we observed the second largest number of results promoting misinformation (63.8%) after Yandex for the English query (71.1%).

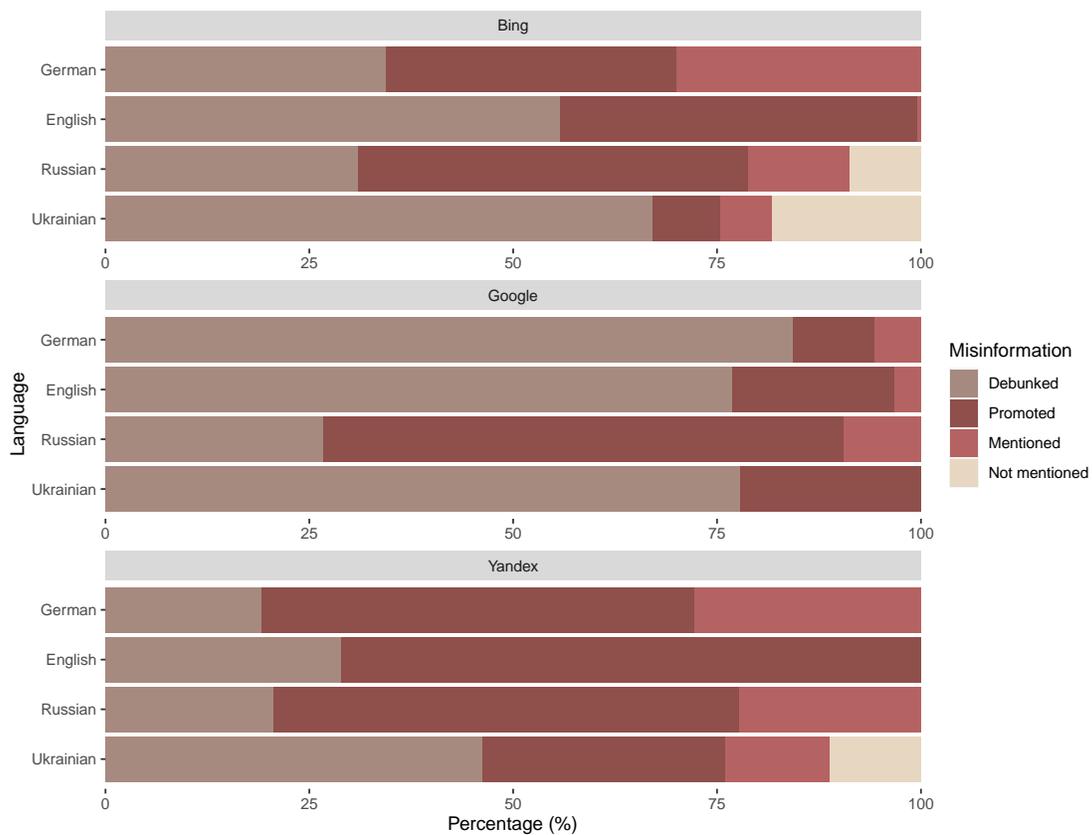

*Figure 3. Misinformation presence by search engine and language of the query*

The results promoting misinformation on Google were most often coming from the Chinese website Russian News and the Russian Duma website. Another source of misinformation was the Ukrainian website for online petitions for the Office of the President of Ukraine. Presumably, the website is intended to be used by Ukrainian citizens for genuine requests, but the broad range of possible authentication mechanisms leave space for provocative petitions, sometimes made by trolls.[3]

Overall, Yandex had a higher proportion of articles promoting misinformation than Google. For all queries, except the Ukrainian one, Yandex showed high proportions of articles promoting misinformation (57.2% for Russian, 53.2% for German, 71.1% for English). Accordingly, results debunking the biolabs narrative were less common than on Google, except for Yandex in Ukrainian, where such articles represented the largest share (46.2%) - but still comparatively lower than Google in Ukrainian (see Figure 3). On Yandex, the sources promoting misinformation that most often featured were Natural News, Great Game India and RT, closely followed by Channel One Russia and Fox News.

On Bing the largest shares of debunking outputs across languages were observed for Ukrainian and English queries (67.1% and 55.7%, respectively). The share of sources that promoted misinformation ranged from 8.3% (Ukrainian) to 47.8% (Russian). Thus, Bing did not prove much more successful than Yandex in debunking the misinformation. The largest shares of articles only mentioning the biolabs narrative (neither promoting nor debunking it), across search engines and languages, were observed for the German query on Bing (30.1%), seconded by German query on Yandex (27.8%). On Bing, the most occuring sources that promoted the biolabs misinformation featured a Youtube video, followed by articles on the French news website Le Courrier des Stratèges and the Swiss Uncut News.

### 4.3. Promoted misinformation in different locations

Figure 4 shows how the exposure to false information across search engines varies depending on the location, the language and time of search queries. When looking solely at the outputs that promote misinformation, we find that for all three locations Google displays a much higher proportion of misinformation for the Russian query (ranging from 60.0% to 80.0%, depending on country and time). In contrast, when using Bing, the proportion of outputs promoting misinformation was higher for the Russian queries, when searching from Germany (from 44.4% to 63.4% depending on the search time) and from France (from 44.3% and 56.9%), whereas, for Switzerland, the highest share of outputs promoting misinformation was in response to the German language query (from 53.0% and 57.8%).

Interestingly, on Yandex the proportion of outputs promoting misinformation was highest in response to the English query (with a maximum of 80.0% in Germany on the morning of the second day and a minimum of 67.1% in France on the morning of the first day). The only case in which we observed a substantial change in the promotion of misinformation over time is for Ukrainian queries on Yandex. Between the morning of 21 June and evening of 22 June, the proportion of content promoting misinformation o dropped from 55.6% to 12.5% for Switzerland, from 55.1% to 4.7% for Germany and from 55.6 to 0.0% for France.

---

[3] Particularly two petitions were prioritized by Google, claiming that "US-funded Biolabs" are "factories of death" and urged to close them down.

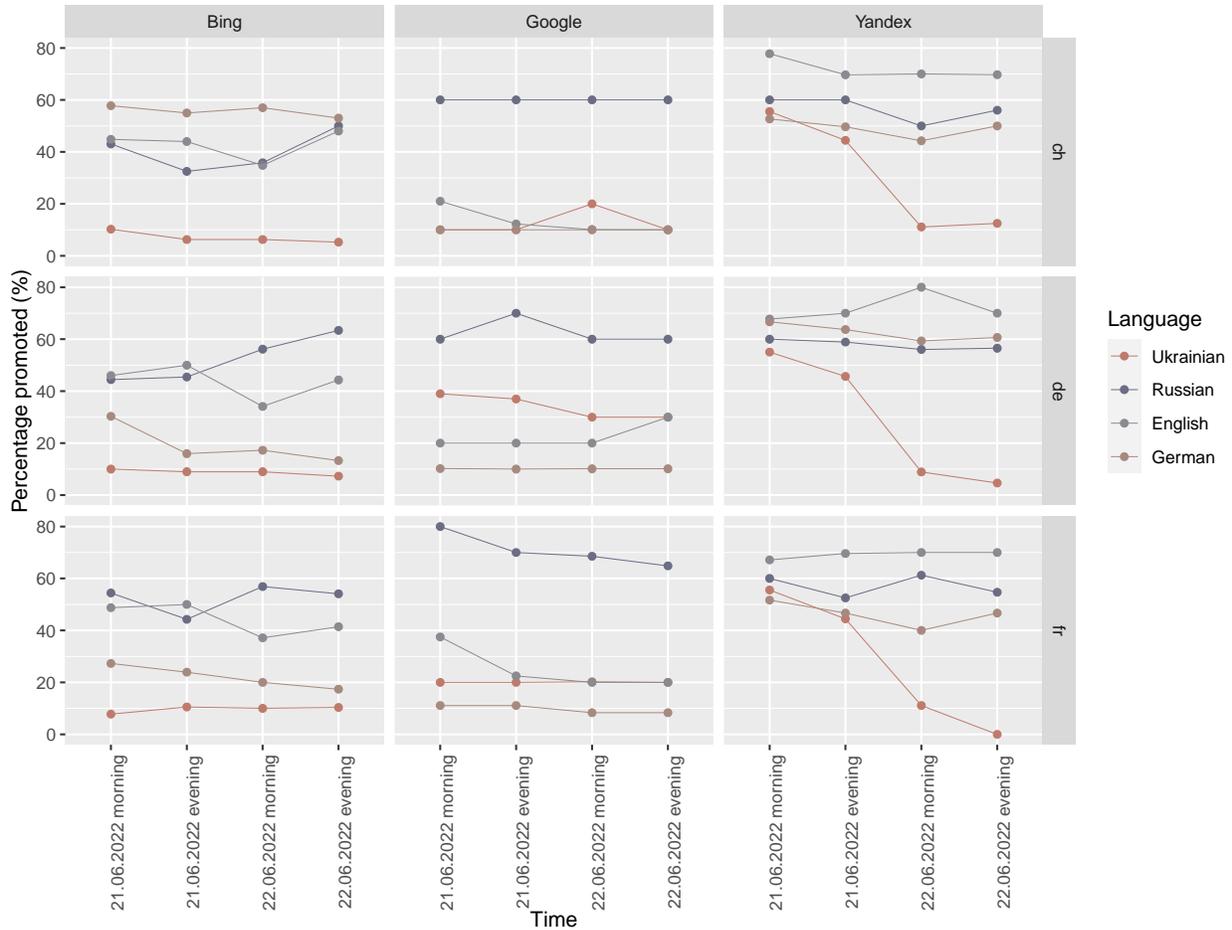

*Figure 4. Proportion of search outputs promoting misinformation by search engine, country of the agent, language of the query, and time*

### 4.4. The effect of variables on the outcome of misinformation

The multinomial regression analysis yields valuable insights into the factors that influence the outcomes of *mentioned* misinformation and *promoted* misinformation[4]. When examining the odds of being exposed to *mentioned* misinformation compared to being exposed to *debunked* misinformation, we observe that conducting a query from Germany does not significantly change the odds compared to conducting the query from Switzerland, other variables being equal. On the other hand, conducting the query from France increases the odds of *mentioned* misinformation exposure by 27% compared to conducting the query from Switzerland.

Language of the query was also associated with the exposure to *mentioned* misinformation compared to *debunked* misinformation: the odds of *mentioned* misinformation for German language queries were 20.60 times those of English language queries. Russian language queries had odds of *mentioned* misinformation 29.45 times the odds of English language queries. Comparatively, Ukrainian language queries had odds of *mentioned misinformation* 3.98 times those of English language queries. As for search engines, Google users had 76% lower odds of being exposed to *mentioned* misinformation compared to Bing, while for Yandex, the odds of such exposure were 2.36% higher than on Bing. The day and time in which the queries were performed were not statistically significantly associated with the odds of *mentioned* misinformation exposure.

---

[4] A summary of the multinomial regression results can be found in Figure 5 with a more detailed breakdown provided in Table 1 in the Supplementary File #1.

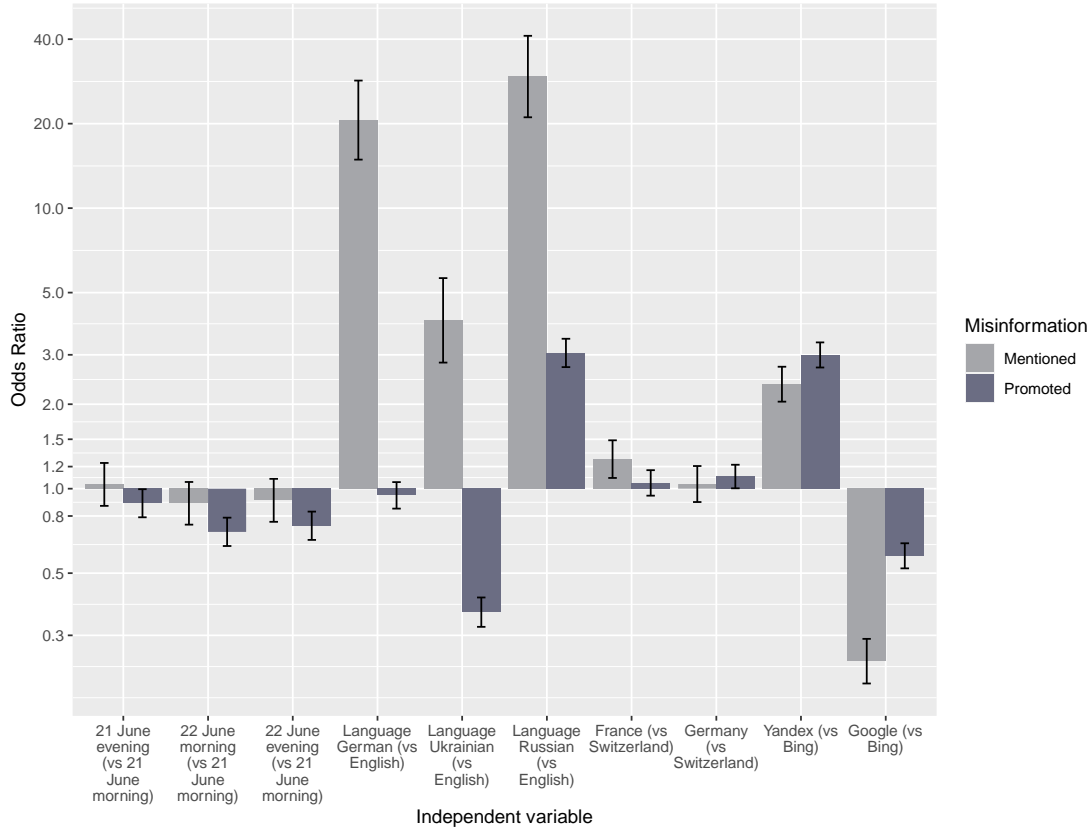

*Figure 5. Odds ratios obtained from the multinomial regression[5].*

As for the odds of being exposed to *promoted* misinformation compared to *debunked* misinformation, then conducting a query from France does not significantly change the odds compared to Switzerland (other variables being equal). On the other hand, conducting the query from Germany increases the odds of *promoted* misinformation exposure by 11% compared to Switzerland. The odds of *promoted* misinformation exposure for Russian language queries were 3.05 times those for English language queries. Ukrainian language queries had lower odds of exposure to *promoted* misinformation compared to English language queries (64% lower), whereas German language queries did not have statistically significant different odds of *promoted* misinformation compared to English queries.

As for search engines, then queries conducted on Google had 42% lower odds of exposure to *promoted* misinformation compared to Bing, while Yandex had 3.0% higher odds of *promoted* misinformation to Bing. Finally, it terms of the time of the search, taking as a reference the morning of June 21, queries performed in the evening of June 21, morning of June 22, and evening of June 22 had lower odds of exposure to *promoted* misinformation (respectively 11%, 30% and 26% lower).

---

[5] 95% confidence intervals are depicted by vertical black segments. Confidence intervals containing the null value (Odds ratio=1) were not considered statistically significant. The y-axis is represented on a logarithmic scale.

## 5. Discussion and Conclusion

In this study, we examined the role of three search engines in the spread of pro-Kremlin misinformation about the US biolabs in Ukraine. For the purpose of analysis, we focused on the top 10 organic search results and identified the type of source from which each search result originates, its political leaning, and whether the source promoted, mentioned or debunked misinformation. Our results show that neither Western search engines nor Yandex fully controlled the dissemination of misinformation concerning the biolabs story. We find that across all languages and locations, the three search engines mentioned or promoted a considerable share of misinformation (33% on Google; 44% on Bing, and 70% on Yandex). Most common misinformation claims related to the extent of US funding for biolabs in Ukraine, the number of US-supported laboratories in Ukraine and the involvement of US elites in the Ukrainian affair. Particularly, users from Germany and Switzerland were more likely to be exposed to search results promoting misinformation.

Firstly, our analysis shows that all search engines in all languages prioritize media outlets in response to the biolabs-related queries. For English and German queries, this often leads to a reduction of misinformation presence in the results. However, for the Russian queries, the media outlets are largely constituted by pro-Kremlin media that results in the opposite effect (i.e. increased exposure to misinformation). This phenomenon stresses the importance of a more thorough accounting for the contextual factors (e.g. the unequal visibility of information sources in specific languages) regarding AICSsto prevent problematic outcomes for the global information environment. Partially, the observed problem aligns with the earlier concerns about the universalization of the AICSs which (partially due to their Western-centric nature) do not necessarily fit the requirements of specific information ecosystems. For instance, the tendency to prioritize the largest journalistic outlets in search engine outputs can be detrimental in the case of searches conducted from the locations or languages associated with the authoritarian regimes due to the largest outlets there often being subjected to state control and promoting the state agenda.

Secondly, the Russian state-influenced search engine Yandex shows, for almost all languages, no presence of independent fact-checking sources, which is another indication of it being censored by the Russian authorities and likely used to circulate more propaganda narratives. This is consistent with the work by (Daucé & Loveluck, 2021) and (Kravets & Toepfl, 2021) who showed that Yandex.News plays an important role in allowing the Kremlin to exert control of the Russian media environment. The only exception to this lack of fact-checking websites on Yandex is represented by the outputs for the Ukrainian query, where a small share of fact-checking websites was retrieved. This points at the lack of curation efforts of results in Ukrainian on Yandex and an overall high number of Ukrainian fact-checkers (International Fact-checking Network Signatories, 2024).

Lastly, the fact that Yandex tends to prioritize Russian narratives is also reflected in the distribution of content by political leaning of the source observed in our study. Yandex and Bing present high shares of Russian state-sponsored and right-wing sources, especially for the Russian language queries. Google is less likely to prioritize such sources, which might be explained by a more selective algorithmic curation model. A direct consequence of these differences is substantial variation in the exposure to pro-Kremlin misinformation that is lower for Google (except queries in Russian) than for Bing and Yandex. The difference between Yandex and Google in terms of source types in response to a query is consistent with the findings of Makhortykh et al. (2022) who found that Yandex tends to prioritize pro-Kremlin information sources, whereas Google is more likely to retrieve outputs criticizing the Russian authorities.

Our study, while comprehensive in its approach, has several limitations. Firstly, the scope of our investigation was confined to the analysis of a single query. While this query was translated into several languages, our findings are not generalizable. Additionally, our study primarily focuses on the short-term changes in curation outcomes. While these insights are valuable, they represent only a snapshot of continuously evolving information ecosystems, which is often the case in algorithm audits due to the evolving nature of platforms' systems (Ballatore, 2015). Lastly, there is a lack of comparable studies in the area of specific Russian misinformation topics, which limits our ability to draw broader conclusions about the distinctiveness of our outcomes. However, our data collection has yielded a more extensive dataset that has not been analyzed in the current study. Thus, in our follow-up research, we aim to contribute to a more nuanced and thorough understanding of how search engines handle and curate content related to misinformation, contributing to our knowledge on the presence of misinformation in search engines' information environments.